\documentclass[12pt]{iopart}
\usepackage[toc,page]{appendix}
\usepackage{placeins} 
\usepackage{subfiles} 
\usepackage[hidelinks]{hyperref}
\usepackage{orcidlink}
\usepackage{color,soul}

\usepackage{cite}

\begin{document}

\hbox{\vspace{-10mm}}
\title{Ad-hoc hybrid-heterogeneous metropolitan-range quantum key distribution network}

\author{
Matthias Goy\textsuperscript{1\textdagger}\,\orcidlink{0009-0006-2330-7226},
Jan Krause\textsuperscript{2\textdagger}\,\orcidlink{0000-0002-3428-7025},
Ömer Bayraktar\textsuperscript{3,4\textdagger}\,\orcidlink{0000-0001-6984-8622},
Philippe Ancsin\textsuperscript{1},
Florian David\textsuperscript{5},
Thomas Dirmeier\textsuperscript{3,4}\,\orcidlink{0000-0002-4935-9810},
Nico Doell\textsuperscript{1},
Jansen Dwan\textsuperscript{1},
Friederike Fohlmeister\textsuperscript{5}\,\orcidlink{0000-0002-2568-7473},
Ronald Freund\textsuperscript{2}\,\orcidlink{0000-0001-9427-3437},
Thorsten A. Goebel\textsuperscript{1},
Jonas Hilt\textsuperscript{2},
Kevin Jaksch\textsuperscript{3,4}\,\orcidlink{0000-0001-9520-0424},
Oskar Kohout\textsuperscript{1},
Teresa Kopf\textsuperscript{1},
Andrej Krzic\textsuperscript{1}\,\orcidlink{0000-0003-2775-4657},
Markus Leipe\textsuperscript{1}\,\orcidlink{0009-0001-3820-5541},
Gerd Leuchs\textsuperscript{3}\,\orcidlink{0000-0003-1967-2766},
Christoph Marquardt\textsuperscript{3,4}\,\orcidlink{0000-0002-5045-513X},
Karen L. Mendez\textsuperscript{1},
Anja Milde\textsuperscript{1},
Sarika Mishra\textsuperscript{1},
Florian Moll\textsuperscript{5},
Karolina Paciorek\textsuperscript{1},
Natasa Pavlovic\textsuperscript{1},
Stefan Richter\textsuperscript{3,4}\,\orcidlink{0000-0003-2128-9272},
Markus Rothe\textsuperscript{1},
René Rüddenklau\textsuperscript{5}\,\orcidlink{0000-0003-2876-8591},
Gregor Sauer\textsuperscript{1},
Martin Schell\textsuperscript{2},
Jan Schreck\textsuperscript{3,4}\,\orcidlink{0009-0003-1651-7354},
Andy Schreier\textsuperscript{2}\,\orcidlink{0000-0002-8424-3899},
Sakshi Sharma\textsuperscript{1},
Simon Spier\textsuperscript{5}\,\orcidlink{0009-0009-0987-7937},
Christopher Spiess\textsuperscript{1}\,\orcidlink{0000-0002-5974-9177},
Fabian Steinlechner\textsuperscript{1,6}\,\orcidlink{0000-0003-0122-1182},
Andreas Tünnermann\textsuperscript{1,6}\,\orcidlink{0000-0003-4018-7626},
Hüseyin Vural\textsuperscript{3,4}\,\orcidlink{0000-0002-5500-5023},
Nino Walenta\textsuperscript{2}\,\orcidlink{0000-0001-7243-0454},
Stefan Weide\textsuperscript{2}
}

\address{\textsuperscript{1}Fraunhofer Institute for Applied Optics and Precision Engineering IOF, Germany}
\address{\textsuperscript{2}Fraunhofer Institute for Telecommunications, Heinrich-Hertz-Institut, HHI, Germany}
\address{\textsuperscript{3}Max-Planck Institute for the Science of Light MPL, Germany}
\address{\textsuperscript{4}Friedrich-Alexander-Universität Erlangen-Nürnberg, Germany}
\address{\textsuperscript{5}German Aerospace Center DLR, Institute of Communications and Navigation, Germany}
\address{\textsuperscript{6}Institute of Applied Physics, Friedrich Schiller University Jena, Germany}
\eads{\mailto{matthias.goy@iof.fraunhofer.de}, \mailto{jan.krause@hhi.fraunhofer.de}}
\address{\textsuperscript{\textdagger}These authors contributed equally to this work.}

\begin{abstract}
This paper presents the development and implementation of a versatile ad-hoc metropolitan-range Quantum Key Distribution (QKD) network. The approach presented integrates various types of physical channels and QKD protocols, and a mix of trusted and untrusted nodes. Unlike conventional QKD networks that predominantly depend on either fiber-based or free-space optical (FSO) links, the testbed presented amalgamates FSO and fiber-based links, thereby overcoming some inherent limitations. Various network deployment strategies have been considered, including permanent infrastructure and provisional ad-hoc links to eradicate coverage gaps. Furthermore, the ability to rapidly establish a network using portable FSO terminals and to investigate diverse link topologies is demonstrated. The study also showcases the successful establishment of a quantum-secured link to a cloud server.
\end{abstract}


\maketitle

\section{\label{sec:introduction}Introduction}

Quantum key distribution (QKD)
\cite{ 
    bennettQuantumCryptographyPublic1984, 
    ekertQuantumCryptographyBased1991
} 
is used to create shared symmetric secret keys between distant parties with information-theoretically provable protocol security
\cite{ 
    gottesmanSecurityQuantumKey2004, 
    rennerSecurityQuantumKey2005, 
    pirandolaAdvancesQuantumCryptography2020 
}.
In contrast, classical cryptographic methods rely on the unproven computational hardness of certain mathematical problems and are hence threatened by the looming of quantum computers yet unknown limitations in computing power.
Thus, 'harvest-now, decrypt-later' attacks
\cite{MigrationPostQuantenKryptografieHandlungsempfehlungen2020}
pose a fundamental thread to the long-term confidentiality of data encrypted today.
In constrast, QKD does not require any assumptions about an adversary's computational power.


Experimental demonstrations of point-to-point QKD links have seen a substantial performance increase over the course of the last decades with systems covering link distances of up to 1002~km
\cite{ 
    korzhProvablySecurePractical2015, 
    boaronSecureQuantumKey2018, 
    lucamariniOvercomingRateDistance2018, 
    wangTwinfieldQuantumKey2022, 
    liu1002KmTwinfield2023}, 
paving the way for nation-wide QKD backbone links.
Massive parallelization
\cite{terhaarUltrafastQuantumKey2023} 
as well as high-performance detectors and high-throughput post-processing
\cite{ 
    walentaFastVersatileQuantum2014,
    grunenfelderFastSinglephotonDetectors2023, 
    liHighrateQuantumKey2023} 
have led to secret-key rates (SKRs) exceeding 100\,Mbit/s over short-range fiber links, establishing information-theoretically secure encryption as a viable option for many applications.

In various situations, establishing direct fiber links can prove impractical in real-world deployments.
Consequently, QKD adapted for free-space optical (FSO) links has been demonstrated with indoor handheld devices
\cite{ 
    vest2014design,
    chun2017handheld,
    vest2022quantum,
    schreier2023beam},
via fiber-wireless-fiber (FWF) terminals
\cite{
    schreier2023coexistence,
    vedovatoRealizationIntermodalFiber2023
    },
unmanned aerial vehicles
\cite{nauerth2013air,quintana2019low},
free-space-coupled terrestrial links over up to 144\,km
\cite{
    steinlechnerDistributionHighdimensionalEntanglement2017, 
    gongFreespaceQuantumKey2018, 
    mollLinkTechnologyAlloptical2022, 
    krzicMetropolitanFreespaceQuantum2023, 
    vedovatoRealizationIntermodalFiber2023}
and satellite-to-ground links
\cite{schmittManderbachExperimentalDemonstrationFreeSpace2007,  
    wang2013direct, 
    liao2017satellite,
    liaoSatelliteRelayedIntercontinentalQuantum2018,
    wang2020satellite, 
    mazzarella2020quarc 
    }.
Further experiments demonstrated the feasibility of multi-protocol transmission
\cite{castillo2023multiprotocol}
as well as QKD access networks solely in free space
\cite{schreier2023building}.
However, impairments due to ambient light
\cite{mollLinkTechnologyAlloptical2022,schreier2023building},
atmospheric turbulence
\cite{castillo2022lab}
and limitations imposed by tracking / steering systems in terms of accuracy and latency
\cite{mollLinkTechnologyAlloptical2022,vest2022quantum,schreier2023coexistence}
must be considered during the design process of a link.

Research also focused on the networking aspects of QKD, investigating ways to integrate QKD into existing network infrastructures
\cite{
    eraerdsQuantumKeyDistribution2010,
    choiQuantumInformationHome2011, 
    kawaharaEffectSpontaneousRaman2011,
    wangLongdistanceCopropagationQuantum2017,
    gavignetCopropagationTb602023},
approaches for scaling QKD networks to more complex network topologies
\cite{
    elliottQuantumCryptographyPractice2003, 
    peevSECOQCQuantumKey2009, 
    chenMetropolitanAllpassIntercity2010, 
    sasakiFieldTestQuantum2011, 
    stuckiLongtermPerformanceSwissQuantum2011, 
    frohlichQuantumAccessNetwork2013, 
    wangFieldLongtermDemonstration2014, 
    joshiTrustedNodeFree2020}, 
the inclusion of end-users
\cite{
    choiQuantumInformationHome2011, 
    frohlichQuantumAccessNetwork2013},
the long-term operation of such networks 
\cite{
    stuckiLongtermPerformanceSwissQuantum2011,
    wangFieldLongtermDemonstration2014
} 
and the integration of QKD in software-defined network architectures allowing for dynamic routing
\cite{
    aguado2017secure,
    aguado2019engineering,
    mehic2020quantum,
    martin2023madqci
}.

This paper presents an overview of various experiments showcasing the feasibility of rapidly established and flexible heterogeneous ad-hoc QKD networks spanning over metropolitan ranges within the Jena QKD testbed.
Expanding on prior experiments conducted between two federal governmental institutions in Bonn in 2021 (cf. Appendix~\ref{app:bonn-demo}), a diverse range of QKD protocols, encompassing entanglement-based, high-dimensional, continuous-variable, and time-bin BB84 systems were employed.
Keys were generated using both direct fiber and FSO links, employing transportable FSO terminals for enhanced mobility and ad-hoc bridging of optical-fiber gaps, e.g. caused by natural disasters.
Additionally, the effectiveness of hybrid links, i.e. FWF links and trusted-node configurations is demonstrated.
The generated keys were securely managed by a key-management layer within our network.
A real-world scenario is emulated by establishing a quantum-secured link to a cloud server hosted wihtin the network.
This paper aims to contribute to the advancement of QKD networks by addressing key challenges in establishing a flexible and secure metropolitan-range ad-hoc network.

This paper provides a comprehensive overview of all experiments conducted within this testbed. Detailed descriptions of these experiments, however, will be individually published by the respective project partners. This approach aims to give a thorough insight into the efforts and collaborations that led to this innovative and efficient ad-hoc metropolitan-range QKD network.

The paper is structured as follows. The logical and cryptographic architecture of the testbed is introduced in Sec.~\ref{sec:architecture}, followed by an in-depth description of the employed FSO terminals and individual QKD systems in Sec.~\ref{sec:subsystems}.
Sec.~\ref{sec:experiment} provides in-depth descriptions of the conducted experiements and implemented application scenarios.
Conclusions are presented in Sec.~\ref{sec:discussion}.

\section{\label{sec:architecture}Architecture}

\begin{figure*}[t]
    \includegraphics[width=\linewidth]{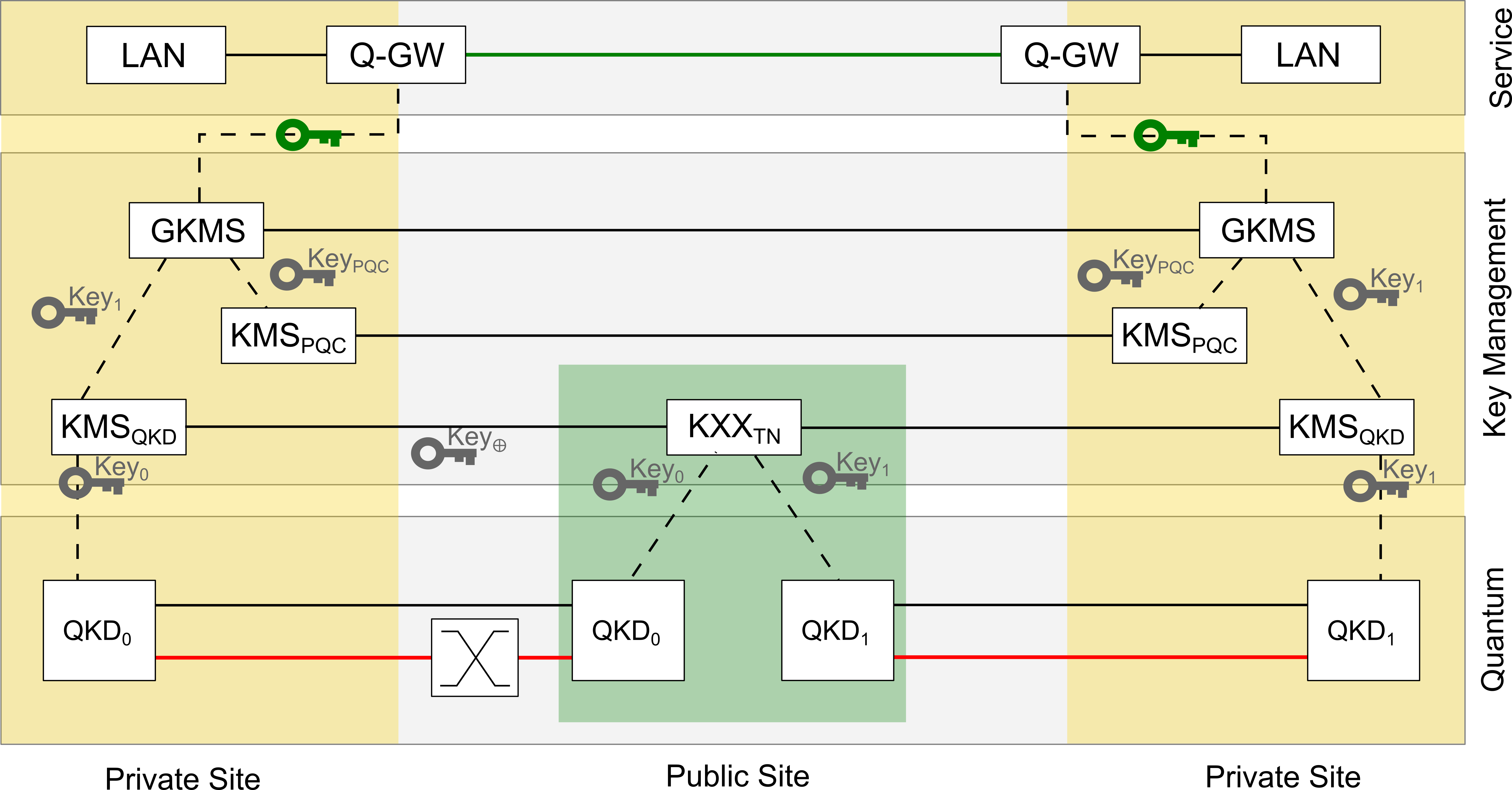}
    \caption{System architecture underlying the following experiments.
    Red background: Trusted node;
    Dashed lines: Path of cryptographic keys;
    Solid lines: Authenticated communication (black), quantum channel (red), quantum-secure communication (green);
    LAN: Local area network;
    Q-GW: Quantum gateway;
    KMS: Key management system;
    GKMS: Global KMS;
    $\mathrm{KMS}_\mathrm{PQC}$: KMS for keys produced by post-quantum cryptography algorithms;
    $\mathrm{KMS}_\mathrm{QKD}$: KMS for QKD keys;
    $\mathrm{KXX}_\mathrm{TN}$: Trusted node key relay;
    QKD$_i$: Quantum key distribution node $i$.
    \label{fig:qkd-i-3}
    }
\end{figure*}

As depicted in Figure~\ref{fig:qkd-i-3}, the network architecture underlying our demonstration experiments follows the ITU-T Y.3800 recommendation~\cite{ITU-T_Y.3800v1.0}.
The \emph{service layer} implements \emph{quantum-secure gateways} (Q-GW), which act as boundaries for the classical communication such that information exchanged between \emph{quantum-secure gateways} is quantum-secure.
In the \emph{key management layer}, all functionalities regarding the management, cryptographic combination and relay of cryptographic keys from several sources are implemented.
Opto-electronic systems for QKD reside in the \emph{quantum layer}.

Our system architecture was specifically designed to take key relays (trusted nodes) into account to also allow for key generation over links with prohibitively large channel losses, cf. Sec.~\ref{sec:architecture:level-2} and Sec.~\ref{sec:experiment:trusted-node}.

In the following, the layers of the system architecture are described in detail.

\subsection{\label{sec:architecture:level-1}Service Layer - Quantum-Secure Gateway}

Commonly, in network architectures, specialized devices are employed as secure gateways to transparently allow secure communication between private sites (red networks) over public networks (black networks) by encrypting, authenticating and ensuring integrity of communication.
The private sites may form a complex network and support various applications.
A secure gateway consumes cryptographic keys generated through out-of-band or in-band channels. 

This paper refers to secure gateways that utilize cryptographic keys, generated through "quantum-secure" methods, as \emph{quantum-secure gateways} (Q-GW).
Given the ongoing standardization and maturation of QKD, the proposed system architecture uses secure gateways that are equipped with standardized interfaces.
These interfaces are designed to accept cryptographic keys exchanged via out-of-band channels, in alignment with the principles of crypto-agility.

\subsection{\label{sec:architecture:level-2}Key Management Layer}

The key management layer acts as an interface between the out-of-band key generation methods and the Q-GW.
This gateway provides symmetric cryptographic keys to the Q-GW via a global key management system (GKMS) upon request.

To allow for trusted nodes in our network, the key management layer also implements a key relay functionality that provides identical keys to the endpoints by propagating Bob's key through the network to Alice using information-theoretically secure XOR operations with keys from the intermediate links.

To enhance the architecture modularity, the implementation of the key management layer uses standardized interfaces.
This also allows all QKD nodes (QKD$_i$) to comprise one or more QKD systems.
By combining keys of several nodes, the system architecture adheres to the principles of crypto-agility.

\subsection{\label{sec:architecture:level-3}Quantum Layer - Quantum Key Distribution}

QKD systems are located within the quantum layer.
Therefore, the QKD nodes (QKD$_i$) require at least two logical channels: an optical channel for the distribution of quantum states (called \textit{quantum channel}) and an authenticated communication channel.
In some cases, systems may also require an auxiliary optical channel, e.g. for synchronization.
All channels might share the same physical medium, assuming precautions are taken against detrimental cross-talk effects.

\section{\label{sec:subsystems}Subsystems}

All subsystems are described in the following subsections and are listed in Tables\,\ref{tab:Subsystems1} and \ref{tab:Subsystems2}.

\subsection{\label{sec:subsystems:fso-terminals}FSO terminals for ad-hoc urban links}

The FSO links discussed in this study were implemented using three distinct types of FSO terminals: two transportable FSO terminals (TFT-1, TFT-2), and the QuBUS, which serves as a transportable transceiver platform for FSO links in QKD networks. All three systems - TFT-1, TFT-2, and the QuBUS - are capable of functioning as both transmitters and receivers. No specialized training is required for their operation, and they can be ready for use within a few hours.

\subsubsection{\label{sec:subsystems:fso-terminals:alice}Transportable FSO terminal TFT-1}

\begin{figure}
    \centering
    \includegraphics[width=0.7\columnwidth]{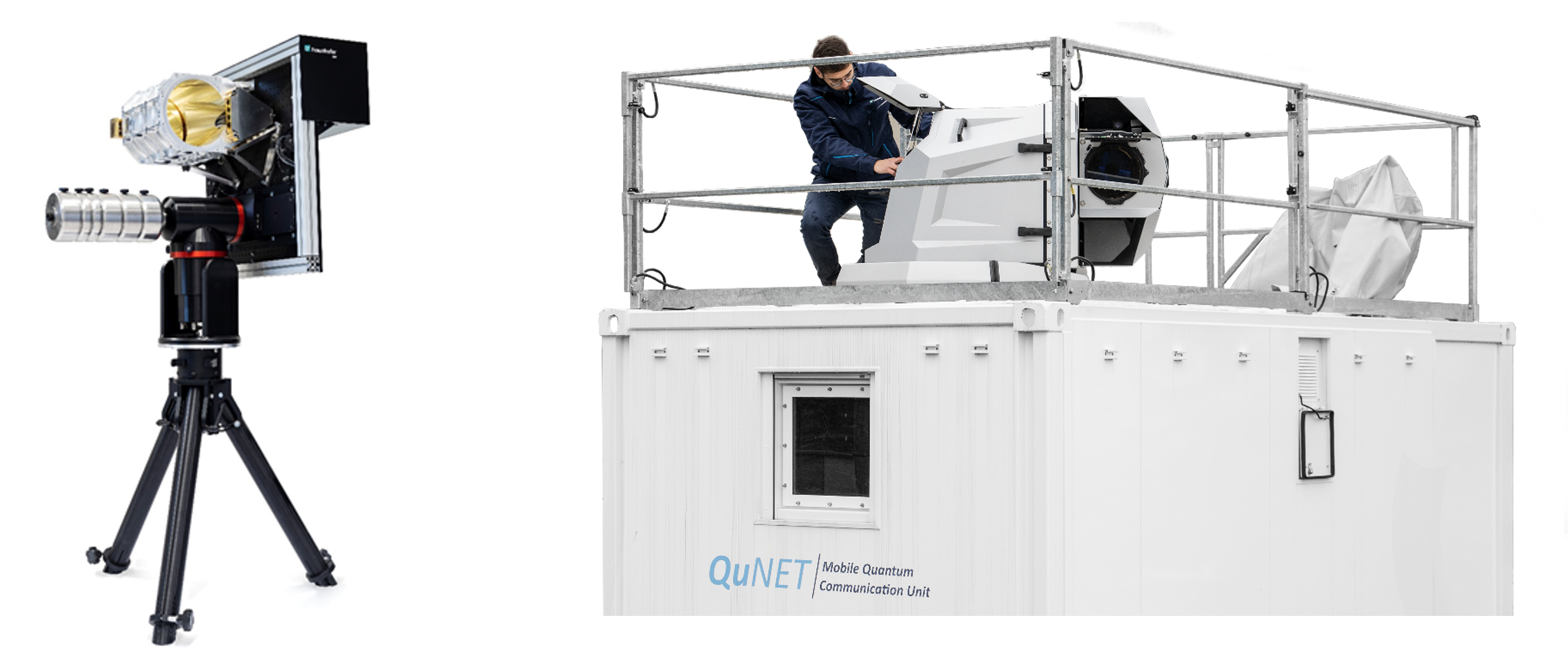}
    \caption{Left: Transportable FSO terminal 1 (TFT-1) mounted on a motorized telescope tripod. Right: QuBUS with periscope assembly on its roof (cf. Sec.~\ref{sec:subsystems:fso-terminals:QuBUS}).}
    \label{fig:TFT-1_QuBUS}
\end{figure}

Fraunhofer IOF developed multi-wavelength transceiver terminals using obscuration-free, diffraction-limited mirror telescopes.
They were designed and assembled as platforms for a broad range of experiments in classical laser and quantum communications.
For the experiments in this paper, the transportable FSO terminal TFT-1
\cite{
    goyHighPerformanceFSOLinks2021,
    krzicMetropolitanFreespaceQuantum2023}
was utilized as transmitter and receiver since it emphasizes the character of ad-hoc metropolitan links that is addressed here.
It uses a 200\,mm aperture telescope, consisting of four metal-based off-axis mirrors.
Mirrors and structural body of TFT-1 are made of an Al6061 alloy with a protected gold coating on all mirror surfaces.
The entire telescope features a magnification of 20, an aperture of 200\,mm and a field of view (FOV) of 3.5\,mrad, and is compactly folded into a volume of ${42 \times 36 \times 26 \, \mathrm{cm}^3}$.
TFT-1 is assembled as a transportable terminal, but, as all its elements are mounted onto a threaded breadboard, it still provides the capability of rearranging optics for versatile experimental use, even allowing for the attachment and free-space coupling of sources and detection modules, cf. Figure ~\ref{fig:TFT-1_QuBUS}.

\begin{table}[!ht]
\caption{\label{tab:Subsystems1}Channel infrastructure utilized during the experiments. A graphical visualization of the channels is given by Figure.\,\ref{fig:geography}.}
\footnotesize
\begin{tabular*}{\textwidth}{p{0.20\textwidth}p{0.7\textwidth}}
\br
Identifier & Subsystem description \\
\mr
TFT-1              & Transportable FSO terminal 1; cf. Sec.~\ref{sec:subsystems:fso-terminals:alice}\\
TFT-2              & Transportable FSO terminal 2; cf. Sec.~\ref{sec:subsystems:fso-terminals:alice}  \\
QuBUS              & Transportable Transceiver terminal; lab container with pointing periscope; cf. Sec.~\ref{sec:subsystems:fso-terminals:QuBUS}  \\
FSO-Link-1         & FSO link between QuBUS and lab container at Stadtwerke Jena; Distance: 1660\,m  \\
FSO-Link-2         & FSO link between a laboratory at Beutenberg Campus and lab container at STW; Distance: 1710\,m  \\
FIBER-Link-1       & Fiber link between QuBUS and laboratory Abbe Center of Photonics (ACP) at Beutenberg Campus; Distance: 300\,m  \\
FIBER-Link-2       & Fiber link between two laboratories at Beutenberg Campus (IOF and ACP); Distance: 685\,m \\
FIBER-Link-3       & Fiber link between the FSO terminal and the QKD systems at STW; Distance: 20\,m \\
\br
\end{tabular*}
\normalsize
\end{table}

\begin{table}[!ht]
\caption{\label{tab:Subsystems2}QKD systems deployed during the experiments.
}
\footnotesize
\begin{tabular*}{\textwidth}{p{0.20\textwidth}p{0.7\textwidth}}
\br
Identifier & QKD system description \\
\mr

BB84-QKD       & Real-time autonomous prepare-and-measure BB84 discrete-variable QKD system (DV-QKD), used with the FWF link; 1550\,nm; cf. Sec.~\ref{sec:subsystems:qkd:hhi-1550-bb84}  \\
BBM92-QKD      & Entanglement-based QKD system; 1550\,nm; cf. Sec.~\ref{sec:subsystems:qkd:iof-1550-bbm92}  \\
HD-QKD         & High-dimensional prepare-and-measure QKD system; 1550\,nm; cf. Sec.~\ref{sec:subsystems:qkd:iof-hd-qkd}  \\
CV-QKD-810     & Continuous-variable QKD system (CV-QKD), free-space-coupled to transmitter and receiver terminals; 810\,nm; cf. Sec.~\ref{sec:subsystems:qkd:mpl-810-cv} L \\
CV-QKD-1550    & CV-QKD system, fiber-coupled to transmitter and receiver terminals; 1550\,nm; cf. Sec.~\ref{sec:subsystems:qkd:mpl-1550-cv}  \\
\br
\end{tabular*}
\normalsize
\end{table}

The terminal is being aligned with the optical path (beacon) of the communication partner by using a visual camera and an iterative adjustment of the coarse (CPA) and fine pointing assemblies (FPA).
After the system is aligned, a control loop takes over and corrects for turbulence-induced fluctuations of position and angle of the incoming beacon and QKD signals.

Its beam stabilization employs a beacon laser at $1064\,\mathrm{nm}$, transmitted by the opposite terminal, and utilizes two position-sensitive devices (PSDs) for analysis.
One PSD is used to measure the position, while another PSD, positioned behind a lens, is used to measure the angle.
To allow for beam stabilization on the opposite terminal, a beacon at the same wavelength is also transmitted in the reverse direction.
This configuration enables the detection of both the beam angle of arrival and the lateral offset with regard to the telescope's pupil.
After correcting these using the fast steering mirrors, a coupling into a single mode fiber can be realized.

\subsubsection{\label{sec:subsystems:fso-terminals:alice2}Transportable FSO terminal TFT-2}

\begin{figure}
    \centering
    \includegraphics[width=1\columnwidth]{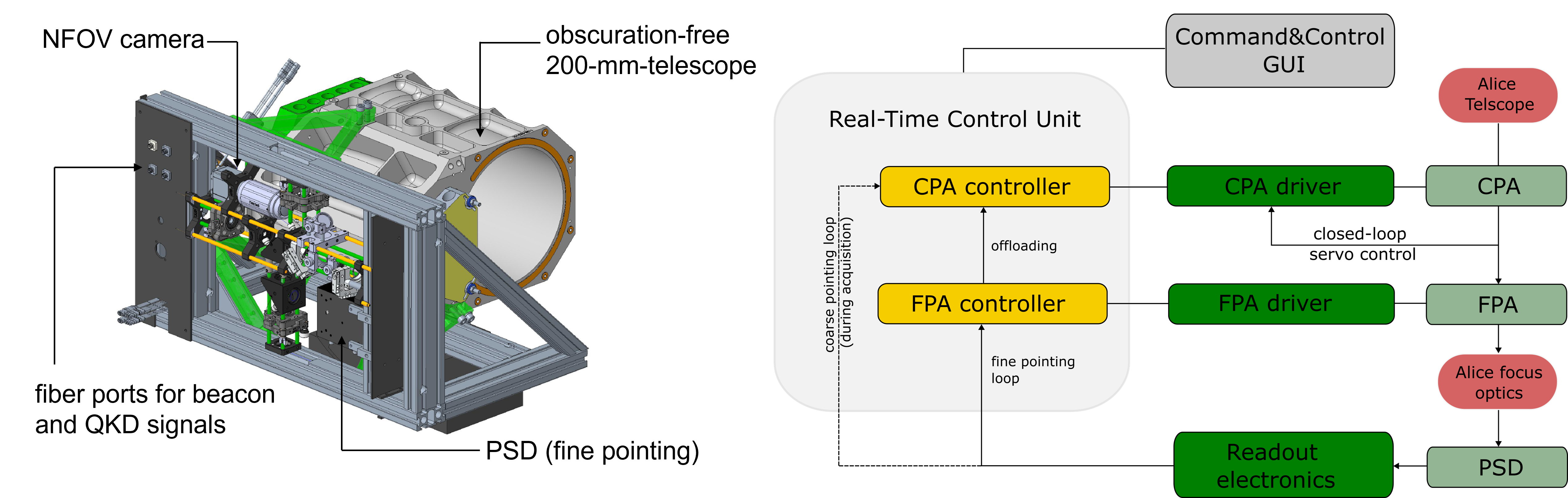}
    \caption{Left: Design of the transportable FSO terminal 2 (TFT-2). Right: Architecture of the pointing, acquisition ant tracking (PAT) system for TFT-2. The PAT system consists of multiple control loops running in parallel. The coarse pointing assembly (CPA) driver features an internal servo control for disturbance rejection. The position-sensitive device (PSD) is used to close the optical loop for the coarse alignment of the CPA to the fine-pointing assembly (FPA) during acquisition and for the fine pointing loop during the tracking phase. An additional integral controller offloads the low-passed FPA angles to the CPA, whenever a predefined threshold is exceeded.
    NFOV: Narrow field of view;
    GUI: Graphical user interface.
    \label{fig:control_loop_Alice_QuNET_beta}}
\end{figure}

In contrast to TFT-1, TFT-2 is assembled more compactly into an aluminum frame, requiring subsystems such as sources and detectors to be connected via fiber.
The TFT-2 telescope features identical mechanical parameters to the TFT-1 telescope.
However, it utilizes a more sophisticated material combination of an aluminum alloy, along with a nickel-phosphorous plating as the base material for the mirrors and the telescope body.
Together with a protected silver coating on its optical surfaces it ensures a higher and more stable optical throughput. 
In addition, TFT-2 is equipped with a control system by DLR that provides an automated alignment and tracking procedure.
The task of the control system is to combine the individual actuators and sensors of CPA and FPA, and thus, suppress the jitter caused by external disturbances, such as vibrations, temperature-induced movements and wind.
For this reason, TFT-2 consists of a CPA with included encoders, a fine steering mirror (FSM) that represents the FPA, and a PSD to detect a reference beacon.
This beacon is transmitted by the opposite terminal (receiver) and is used for offloading the FPA and CPA of the transmitter.
This functionality is implemented on a third-party proprietary real-time system (Speedgoat\textregistered Unit real-time target machine), specifically developed for the execution of control loops.
The pointing, acquisition, and tracking (PAT) system implementation is sketched in Figure~\ref{fig:control_loop_Alice_QuNET_beta}. 
First, the transmitter is roughly aligned with the target.
Then, a raster or spiral search is started.
As soon as a beacon signal from the opposite terminal is detected, the closed-loop tracking mode is activated.
As the FOV is larger than the actuation range of the FPA, a valid signal indication might occur outside of the FSM range.
For this particular reason and due to the drift described earlier, an offloading mechanism is used.
As soon as the FSM exceeds a threshold value, the CPA compensates for the deflection of the FSM within one full revolution in azimuth and $\pm 15^\circ$ in elevation.
By selecting suitable controller parameters, it is possible to separate slow changing disturbances, such as temperature fluctuations, from fast jitter, as e.g. caused by vibrations.
In that case, the FSM accounts for the suppression of the remaining high frequencies.
The residual jitter is kept low to ensure reasonable losses.
The controller is able to reduce the pointing error even under strong turbulence on the link. The standard deviation for pointing jitter is measured as $338.6\, \mu\mathrm{rad}$ ($80.3\,\mu\mathrm{rad}$) for open- (closed-) loop operation.

\subsubsection{\label{sec:subsystems:fso-terminals:QuBUS}Transportable terminal QuBUS}

The QuBUS is designed as a robust experiment vehicle for ad-hoc field measurement campaigns in quantum communications.
Basis for the QuBUS is a 15-foot shipping container which provides a laboratory infrastructure including an optical table ($1.2 \times 1.5\,\textrm{m}^2$), space for racks, monitors, humidity and temperature control, and connections to a variety of classical wireless and fiber-based networks.
A highly stable, precise periscope is installed on the roof of the QuBUS which enables a coarse pointing within a $‑5^\circ$ to $90^\circ$ angle in elevation and a $360^\circ$ angle in azimuth.
The optical axis of the periscope leads inside the QuBUS towards the optical table below the periscope where TFT-1 is installed.
For the experiments described in this paper, the telescope and beam stabilization system of TFT-1 were used, cf. see Sec.~\ref{sec:subsystems:fso-terminals}.
In this configuration, the periscope represents the CPA whereas a set of two tip/tilt mirrors are representing the FPA.
Although the periscope provides PAT capability, the telescopes for the experiments were manually aligned using an iterative improvement of the optical link transmission.
In addition to the link optics, a weather station and a scintillometer were used to record environmental and atmospheric conditions during the experiments.
For the experiments presented here, the QuBUS was connected to the local campus fiber network with a direct connection to a laboratory.

\subsection{\label{sec:subsystems:qkd}QKD systems}

\subsubsection{\label{sec:subsystems:qkd:hhi-1550-bb84}Time-bin BB84 DV-QKD at \texorpdfstring{1550\,nm}{1550 nm}}

The real-time prepare-and-measure discrete-variable QKD system developed by Fraunhofer HHI (BB84-QKD, cf. Table~\ref{tab:Subsystems2}) utilizes the 1-decoy BB84 QKD protocol
\cite{
    ruscaFinitekeyAnalysis1decoy2018,
    wiesemannConsolidatedAccessibleSecurity2024}.
This system employs time-phase-encoded qubits, where the computational basis $\mathsf{Z}$ consists of single-photon states in either an early or late time-bin, and the superposition basis $\mathsf{X}$ consists of single-photon states in both time-bins with a phase difference of 0 or $\pi$, respectively.

\begin{figure}
    \centering
    \includegraphics[width=\columnwidth]{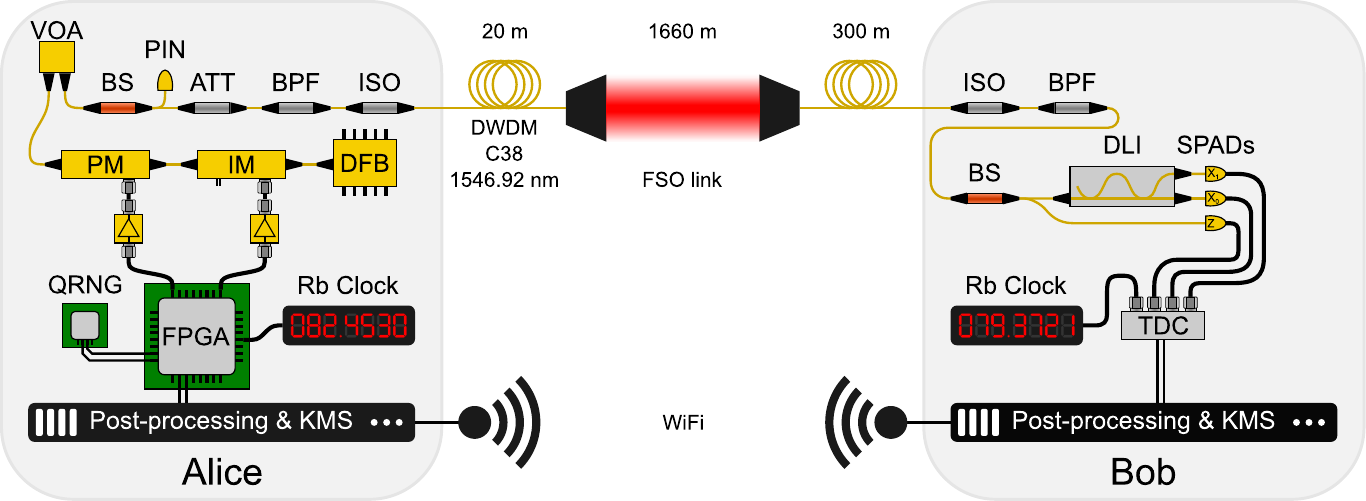}
    \caption{\label{fig:HHI-setup}System design of the 1-decoy time-bin BB84 QKD system from Fraunhofer HHI (BB84-QKD).
    ATT: Fix attenuator;
    BPF: Band-pass filter;
    BS: Beam splitter;
    DFB: Distributed-feedback laser;
    DLI: Delay-line interferometer;
    FPGA: Field-programmable gate array;
    IM: Intensity modulator;
    ISO: Isolator;
    KMS: Key-management system;
    PIN: Photodiode;
    PM: Phase modulator;
    QRNG: Quantum random number generator;
    Rb Clock: Rubidium atomic clock without GPS link;
    SPAD: Single-photon avalanche diode;
    TDC: Time-to-digital converter;
    VOA: Variable optical attenuator.
    }
\end{figure}

At the transmitter (Alice), the qubits are prepared with a frequency of 625\,MHz (two 800\,ps timebins per qubit) by tailoring the optical output of a C-band continuous-wave laser (ITU DWDM C38) using an intensity modulator, a phase modulator, as well as a variable and a fixed optical attenuator to regulate the output photon number per qubit. 
The choice of each qubit state is randomly determined based on the output of a commercial quantum random number generator (QRNG, ID Quantique IDQ20MC1-T), followed by an advanced encryption standard counter mode (AES-CTR) random-number expansion scheme implemented on a field-programmable gate array (FPGA, Xilinx Ultrascale+ VU13P).
This allows for interruption-free continuous operation.

At the receiver (Bob), a fiber coupler splits the incoming states towards a delay-line-interferometer (DLI) with single-photon avalanche diodes (SPADs) in each interferometer output for measurements in the superposition basis, and towards a SPAD for measurements of the arrival time in the computational basis. The passive basis choice is random and is indicated by the detection event at a detector of either path.
For all described experiments the SPADs (ID Quantique IDQube) were operated in free-running mode with a dead-time of 20~$\mu$s and an efficiency of 25~\%.

To protect against Trojan-horse attacks
\cite{gisinTrojanhorseAttacksQuantumkeydistribution2006}
and detector backflash attacks
\cite{kurtsieferBreakdownFlashSilicon2001},
an optical isolator and an optical band-pass filter are incorporated into both the transmitter and receiver systems.
The receiver system employs a time-to-digital converter that is synchronized with the Bob's master clock.

There is no need for an additional clock synchronization channel. This is made possible by the use of two Rubidium atomic clocks that supply the master frequency for each system. A new synchronization technique compensates for residual clock drifts using brief synchronization sequences that are also conveyed over the quantum channel and interleaved with the transmitted qubits \cite{krauseFlexibleRealTimeQuantum2024}. Given that the Rubidium clocks do not need a Global Positioning System (GPS) reference, this strategy offers maximum flexibility and is not reliant on external infrastructure.

The system features a fully-automatic initialization procedure and active feedback loops to continuously maintain low quantum bit error rates (QBER) in both bases.
Thus, single-button operation of the system has been achieved.

For QKD post-processing, a commercial software solution provided by the Austrian Institute of Technology (AIT) is used.
For all experiments a
security parameter $\epsilon_\mathrm{sec} = 10^{-9}$,
correctness parameter $\epsilon_\mathrm{cor} = 10^{-15}$,
mean photon numbers $\mu_\mathrm{signal} = 0.47$ and $\mu_\mathrm{decoy} = 0.17$,
decoy probability $p_\mathrm{decoy} = 0.5$,
and a block size $N = 0.5\times 10^5$ were used.
The distilled secret keys are pushed into the local key management system (LKMS), which provides them to the GKMS using the standardized ETSI GS QKD 004 key interface protocol \cite{ETSIGSQKD_004_2020}.

This QKD system was used for the experiments described in Sec.~\ref{sec:experiment:fwf-dv-qkd} and \ref{sec:experiment:trusted-node}.

\subsubsection{\label{sec:subsystems:qkd:iof-1550-bbm92}BBM92-QKD at \texorpdfstring{1550\,nm}{1550 nm}}

\begin{figure}
    \centering
    \includegraphics[width=152.863 mm]{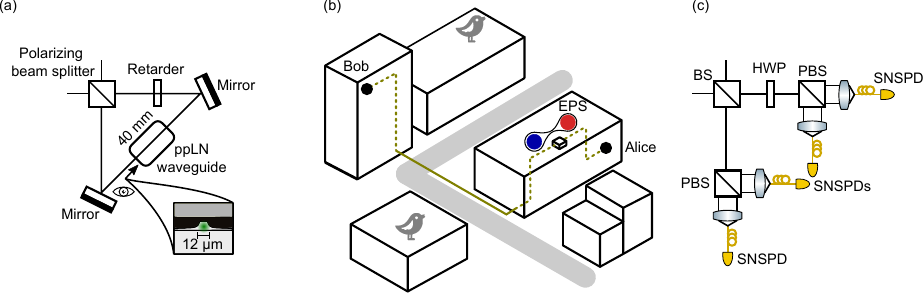}
    \caption{Entanglement-based QKD. (a) The entangled photon pair source (EPS) consists of a ppLN waveguide that is pumped bidirectionally in a Sagnac loop with a 775\,nm continuous-wave laser. Two entangled photons are generated at a center wavelength of 1550\,nm via type-0 spontaneous parametric down-conversion.  
    (b) The EPS transmits the photons to the receiver Alice in the same building and receiver Bob in a neighboring building. (c) The polarization of the light is detected in a polarization analysis module. BS: beam splitter; HWP: half-wave plate; PBS: polarizing beam splitter; SNSPD: superconducting nanowire single photon detector.}\label{fig:iof_QKD_SE1_Entanglement_setup}
\end{figure}

The QKD system developed by Fraunhofer IOF utilizes the BBM92 protocol that is based on entangled photon pairs in the polarization degree of freedom (DoF) (BBM92-QKD, cf. Table~\ref{tab:Subsystems2}).
The entangled photon pairs are generated by the mechanism of spontaneous parametric down-conversion (SPDC) which happens inside the entangled photon pair source (EPS).
The type-0 down-conversion takes place in a periodically-poled Lithium Niobate (ppLN) waveguide of 40\,mm length that is pumped bidirectionally in a Sagnac-loop configuration (Figure~\ref{fig:iof_QKD_SE1_Entanglement_setup}).
The pump at 775\,nm operates in continuous-wave mode and generates photon pairs with the same polarization (type-0 down-conversion) at the central wavelength of 1550\,nm.
The design of the source combines bulk and integrated optics to achieve a high pair generation rate of up to $73.7 \times 10^6$ pairs s$^{-1}$ mW$^{-1}$.
The Sagnac loop configuration allows for the generation of entangled pairs in the polarization DoF and achieves visibilities of $>99.6\,\%$ in both the horizontal/vertical and diagonal/antidiagonal bases. Furthermore, the generated photon pairs are filtered and spectrally demultiplexed at 1530\,nm for the signal and 1570\,nm for the idler photon.

After the photons have been demultiplexed, one is measured locally (Alice) and the other is sent to a remote location (Bob), both equipped with superconducting nanowire single-photon detectors (SNSPD, Single Quantum) with detection efficiencies $>85\,\%$ and timing jitters $<40$\,ps, FWHM. The two receivers, Alice and Bob, have a polarization analysis module to measure the state. Afterwards, the arrival time and state information is processed and a secure key extracted. The QKD post-processing is implemented by a commercial software solution provided by the Austrian Institute of Technology. 

This QKD system was used for the experiment described in Sec.~\ref{sec:experiment:eb-qkd}.

\subsubsection{\label{sec:subsystems:qkd:iof-hd-qkd}High-dimensional time-bin QKD at \texorpdfstring{1550\,nm}{1550 nm}}

\begin{figure}[!b]
    \centering
    \includegraphics[width=94.102mm]{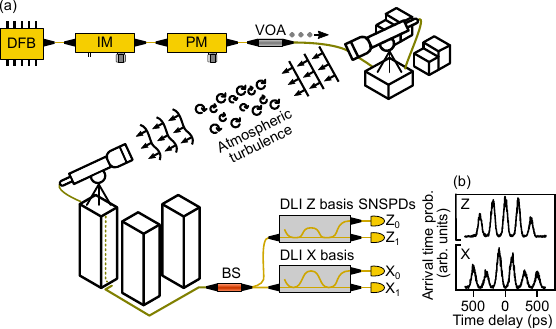}
    \caption{\label{fig:high_D_setup}
     High-dimensional QKD. (a) The prepare-and-measure QKD system was deployed on a 2385\,m hybrid link (1700\,m FSO link + 685\,m fiber link). The essential parts of the QKD system are an intensity and phase modulator at Alice and two interferometers at the receiver Bob. The two interferometers have an imbalance of 200 ($\tau$) and 400\,ps ($2\tau$). The output is measured through SPADs. (b) The arrival time statistics represent 5 distinct time slots in the $\mathsf{Z}$ basis and 6 time slots in the $\mathsf{X}$ basis.
     DFB: Distributed-feedback laser;
     IM: Intensity modulator;
     PM: Phase modulator;
     VOA: Variable optical attenuator;
     BS: Beam splitter;
     DLI: Delay-line interferometer;
     SNSPD: Superconducting nanowire single-photon detector.
    }
\end{figure}

The prepare-and-measure QKD System by Fraunhofer IOF is based on the higher dimensional time-bin 1-decoy state BB84 protocol (HD-QKD, cf. Table~\ref{tab:Subsystems2}).
The system employs 4-dimesional states in the time-phase DoF (Figure \ref{fig:high_D_setup}).
At the transmitter (Alice), a continuous-wave laser with a central wavelength of $1550\,\mathrm{nm}$ (Thorlabs, single mode fiber-pigtailed) and an intensity modulator are used to prepare a train of weak coherent pulses.
The transmitter unit for state preparation consists of an intensity modulator (IM), phase modulator (PM) and variable optical attenuator (VOA) - typically used in BB84 implementations, cf. Sec~\ref{sec:subsystems:qkd:hhi-1550-bb84}.
The modulators are used to prepare states in the $\mathsf{Z}$ and $\mathsf{X}$ basis with probability $P_\mathsf{Z}$ and $P_\mathsf{X}$, respectively.
A subsequent VOA is used to get the single photon level.
Each state prepared in the $\mathsf{Z}$ and $\mathsf{X}$ basis is a superposition of two time bins \cite{vagnilucaEfficientTimeBinEncoding2020}.
The temporal separation between two consecutive time bins $\tau$ is $200\,\mathrm{ps}$ and the states are prepared with a clock rate of $500\,\mathrm{MHz}$ with timing jitter smaller than 50\,ps (FWHM). Every clock cycle has a time buffer to the left and right of the state to simplify the peak recovery.
The probability to select the $\mathsf{Z}$ basis is 80\,\% and is used for key generation. The $\mathsf{X}$ basis serves to estimate the error rate. The prepared state is sent to Bob via the quantum channel.
At the receiver side (Bob), two imbalanced interferometers with a delay of $\tau$ and $2\tau$ are used to measure the $\mathsf{Z}$ and $\mathsf{X}$ basis, respectively.
The output state of interference is measured through SNSPDs with detection efficiencies $>85$\,\% and post-processed through an IOF-developed software in Python.

This QKD system was used for the experiment described in Sec.~\ref{sec:experiment:hd-timebin-qkd}.

\subsubsection{\label{sec:subsystems:qkd:mpl-810-cv}Free-space CV-QKD at \texorpdfstring{810\,nm}{810 nm}}

\begin{figure}
    \centering
    \includegraphics[width=0.9\columnwidth]{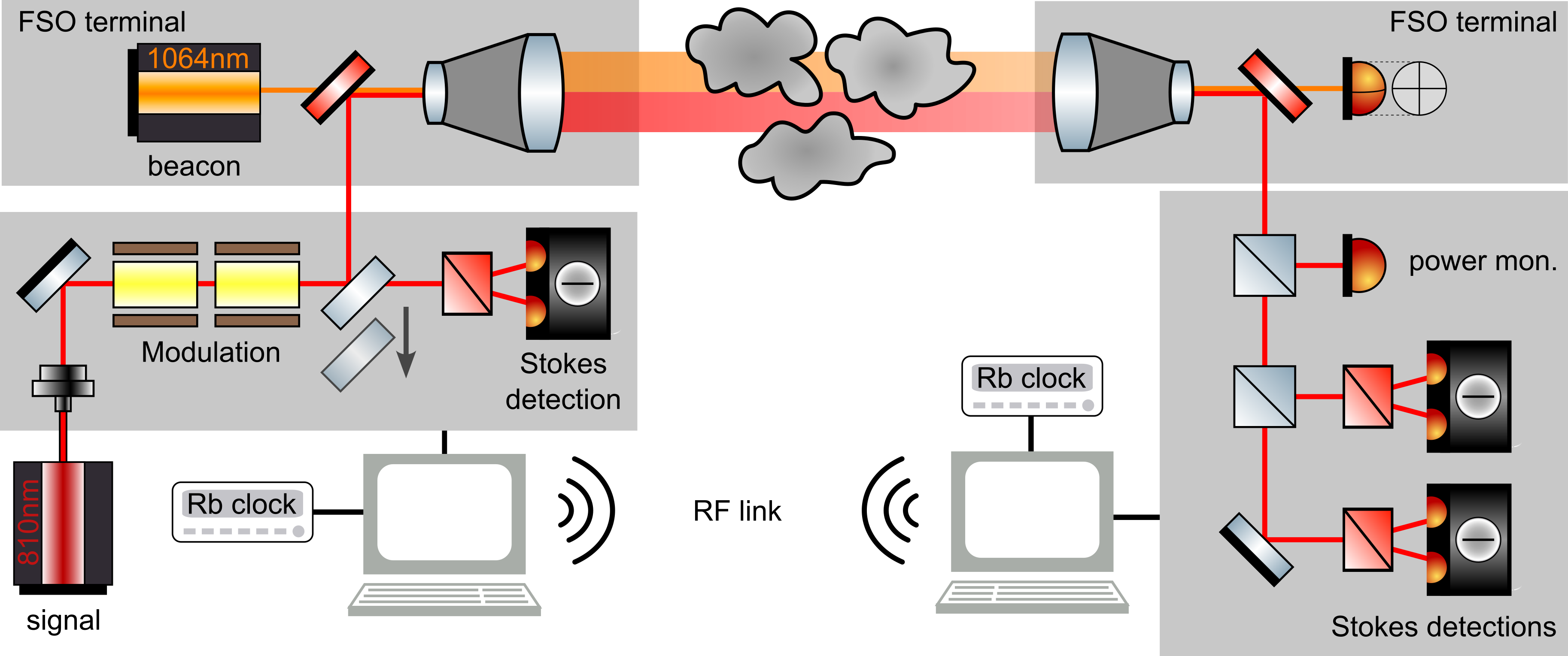}
    \caption{\label{fig:MPL-setup}Schematic of the opto-electronic free-space CV-QKD system from MPL, used in a point-to-point configuration with the FSO terminals provided by IOF.
    Rb: Rubidium;
    RF: Radio-frequency.
    }
\end{figure}

The CV-QKD system by MPL uses polarization encoding with discrete modulation (DM) \cite{heimAtmosphericQuantComm2014,jaksch2024composablefreespacecontinuousvariablequantum} (CV-QKD-810, cf. Table~\ref{tab:Subsystems2} and Figure~\ref{fig:MPL-setup}).
At the sender, a strong and circularly polarized reference beam (local oscillator) with an additional 25\,MHz discrete modulation of four coherent states in the S1 and S2 Stokes parameters is prepared, equivalent to a quadrature phase shift keying (QPSK) modulation in optical phase space.
The local oscillator and the signal states share the same spatial mode and are sent to the QuBUS using TFT-1. As a source of entropy, pre-acquired and locally stored random numbers from a QRNG are used.

At the receiver, a small fraction of the beam is split off to monitor the transmission of the atmospheric link.
The remainder of the beam is then measured by a heterodyne detection scheme realized by the simultaneous measurement of two Stokes parameters using balanced PIN photodetectors with a quantum efficiency of 0.92 and a bandwidth of 65\,MHz.
Since the local oscillator and the signal propagate in the same spatial mode, they experience the same atmospheric wavefront distortions which are then auto-compensated during optical interference. By that, negative influences of atmospheric fluctuations on the visibility can be eliminated. Furthermore, the local oscillator also acts as an intrinsic spatial and narrow-band linewidth filter against straylight, rendering daylight operation feasible.
Clock synchronization between sender and receiver is achieved using two Rubidium atomic clocks.

This QKD system was used for the experiment described in Sec.~\ref{sec:experiment:free-space-cv-qkd}.

\subsubsection{\label{sec:subsystems:qkd:mpl-1550-cv}Fiber-based CV-QKD at \texorpdfstring{1550\,nm}{1550 nm}}

\begin{figure}
    \centering
    \includegraphics[width=0.9\columnwidth]{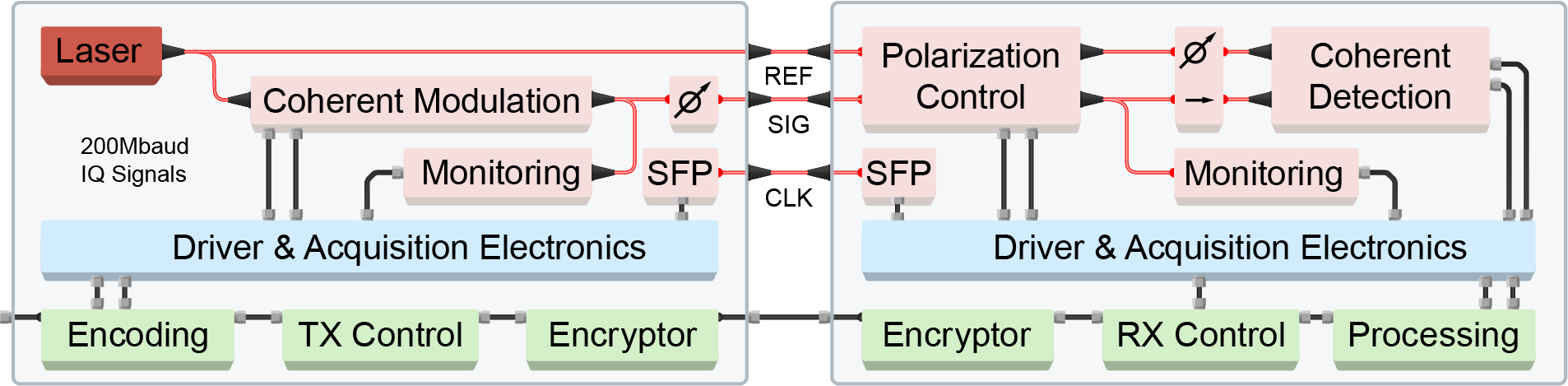}
    \caption{\label{fig:MPL-fiber-setup}Schematic overview of MPL/FAU's 1550\,nm DM-CV-QKD transmitter (left) and receiver modules (right).
    The colored shadings indicate the optical (red), analog (blue) and digital (green) subdomains.
    IQ: In-phase quadrature;
    TX: Transmitter;
    RX: Receiver;
    REF: Optical reference signal;
    SIG: Quantum signal;
    CLK: Clock signal;
    SFP: Small form factor pluggable optical transceiver.
    }
\end{figure}

MPL and FAU also field-tested a fiber-coupled CV-QKD system (CV-QKD-1550, cf. Table~\ref{tab:Subsystems2}) based on multiplexing of coherent states at an optical carrier wavelength of 1544.53\,nm (ITU DWDM C41), cf. Figure \ref{fig:MPL-fiber-setup}.
The sender module houses an ultra-low noise narrow-bandwidth laser source, whose output is asymmetrically split into a bright reference and a signal light path.
A bias-stabilized, nested Mach-Zehnder I/Q modulator driven by a wideband digital-to-analog converter (DAC) imparts a 200\,Mbaud QPSK modulation sideband-shifted by 250\,MHz onto the signal light, interleaved with a fixed pattern of phase reference symbols.
The signal is attenuated to a level of less than one photon per symbol on average, monitored with a photodiode tap-off and coupled alongside the reference light into a duplex fiber cable to the receiver.
If required, the sender may couple light emitted by an SFP (Small form factor pluggable optical transceiver) module into the same path and polarization mode as an auxiliary bright beacon.

The receiver module compensates for polarization drifts in the fiber using two electrically driven polarization controllers.
A set of low-loss optical switches allows selective input blocking required for fully automatic stabilization and noise calibration procedures.
Both inputs interfere in an optical $90^\circ$ hybrid (a specific arrangement of beam splitters and phase shifters), whose outputs are guided to a pair of balanced PIN photodetectors (efficiency: $\eta \approx 0.76$, bandwidth: 500\,MHz). An analog-digital converter acquires the resulting electric signals to measure the I/Q quadratures.
From these measurements, a processing system in authenticated communication with the sender continuously estimates relevant channel parameters such as excess noise and an asymptotic secret-key ratio following \cite{zhangImprovingPerformanceFourstate2012} and \cite{denysExplicitAsymptoticSecret2021}.

This QKD system was used for the experiment described in Sec.~\ref{sec:experiment:fiber-cv-qkd}.

\section{\label{sec:experiment}QKD testbed and experiments}

\begin{figure}[!b]
    \centering
    \includegraphics[width=0.8\linewidth]{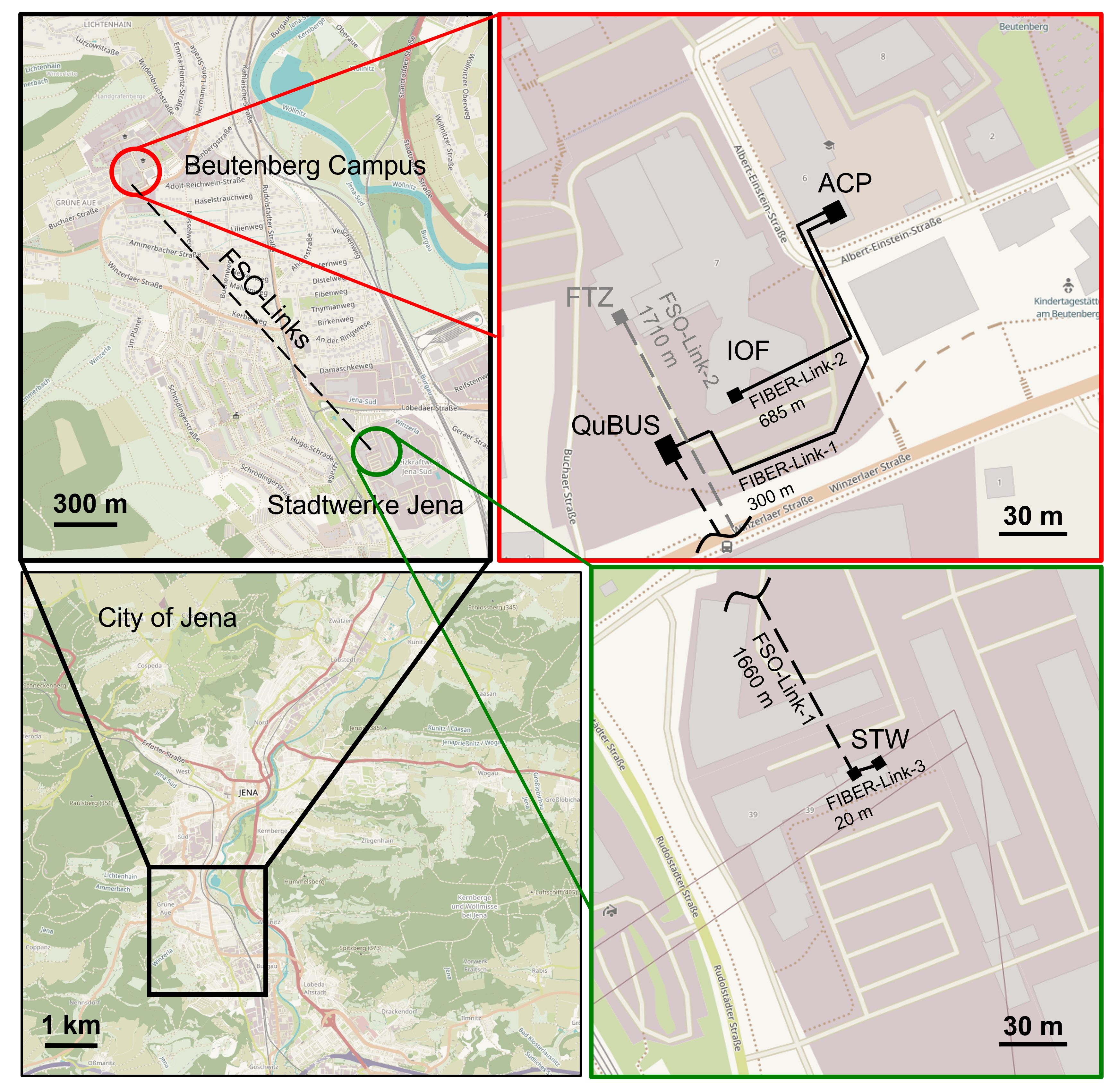}
    \caption{\label{fig:geography}
    The Jena QKD testbed used for the experiments (cf. Table \ref{tab:Subsystems1}).
    Bottom left and top left: Map of Jena with the framed area where the experiments were performed – Beutenberg Campus and Stadtwerke Jena.
    Bottom right and top right: Closer look to Beutenberg Campus (ACP: Abbe Center of Photonics, IOF: Fraunhofer Institute for Applied Optics and Precision Engineering, FTZ: Fiber Technology Center) and Stadtwerke area (STW) with the nodes and links that were used within the experiments.
    }
\end{figure}

Experiments were conducted in the fiber and FSO testbed in Jena, which is comprised of a northern area, housing the Fraunhofer IOF and the Abbe Center of Photonics (ACP), and a southern area, where the local energy supplier Stadtwerke Jena (STW) is located, as depicted in Figure 9. An urban FSO link of 1660m (FSO-Link-1) bridged these two areas, extending between two laboratory containers. The design of the testbeds was intended to replicate a scenario where an ad-hoc FSO interconnect is necessitated between two isolated network segments, potentially in response to a natural disaster scenario.
Direct QKD links were established (cf. Sec. \ref{sec:experiment:hd-timebin-qkd}, \ref{sec:experiment:free-space-cv-qkd}, \ref{sec:experiment:fiber-cv-qkd}, \ref{sec:experiment:eb-qkd}), a Fixed Wireless Fiber (FWF) QKD link was implemented (cf. Sec. \ref{sec:experiment:fwf-dv-qkd}), and a trusted node QKD link was initiated (cf. Sec. \ref{sec:experiment:trusted-node}).

\subsection{\label{sec:experiment:hd-timebin-qkd}Direct link: High-dimensional time-bin QKD}

\begin{figure}[!b]
    \centering
    \includegraphics[width=63.327mm]{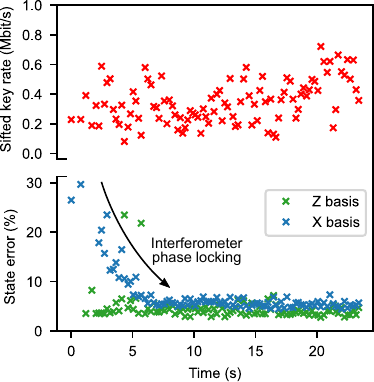}
    \caption{\label{fig:high_D_exp}
    Measurement results for the HD-QKD system.
    The sifted-key rate was approximately 350 kbit/s  The error rate of the states reduces in the first 10 seconds through locking of the interferometer by means of the phase information of the single photons. The mean error rate is 4.1\,\% and 5.5\,\% in the $\mathsf{Z}$ and $\mathsf{X}$ basis, respectively.  
    }
\end{figure}

The high-dimensional time-bin QKD system (HD-QKD, cf. Sec.~\ref{sec:subsystems:qkd:iof-hd-qkd}) was deployed on FSO-Link-1 between STW (Alice) and QuBUS (Bob) demonstrating a 2385\,m hybrid link (1700\,m FSO link + 685\,m fiber link) with atmospheric turbulence (Figure \ref{fig:high_D_exp}(a)).

The probability distribution of the arrival time after the interferometer indicated 5 slots in the $\mathsf{Z}$ basis and 6 slots in the $\mathsf{X}$ basis, cf. Figure \ref{fig:high_D_exp}(b).
A sifted key rate of 350\,kbit/s was obtained, after carefully selecting the interfering time slots for both measurement bases, cf. Figure \ref{fig:high_D_exp}(c).

The sifted-key rate originates in the sum of all detections in $\mathsf{Z}$ basis if the state prepared by Alice was given in the same $\mathsf{Z}$ basis. The error of the detected states reduced in the first seconds of the measurement as the interferometers are locked in phase by processing the arrival time information of the single photons. 

The high-dimensional link overcomes the limit set by the saturation of the detector in the low attenuation regime by increasing the amount of information per bit.
Implementation of the finite-key security analysis together with the one decoy state protocol ensures security of the key transfer \cite{vagnilucaEfficientTimeBinEncoding2020,ruscaFinitekeyAnalysis1decoy2018}. Decoy state intensities and their respective transmission probabilities were chosen based on a previously performed SKR-maximizing parameter optimization. Therefore, the feasibility of 4-dimensional state transmission on turbulent atmospheric links is one step further in the direction of robust and flexible QKD links in deployed scenarios.

\subsection{\label{sec:experiment:free-space-cv-qkd}Direct link: Free-space CV-QKD}

\begin{figure}
    \centering
    \includegraphics[width=\linewidth]{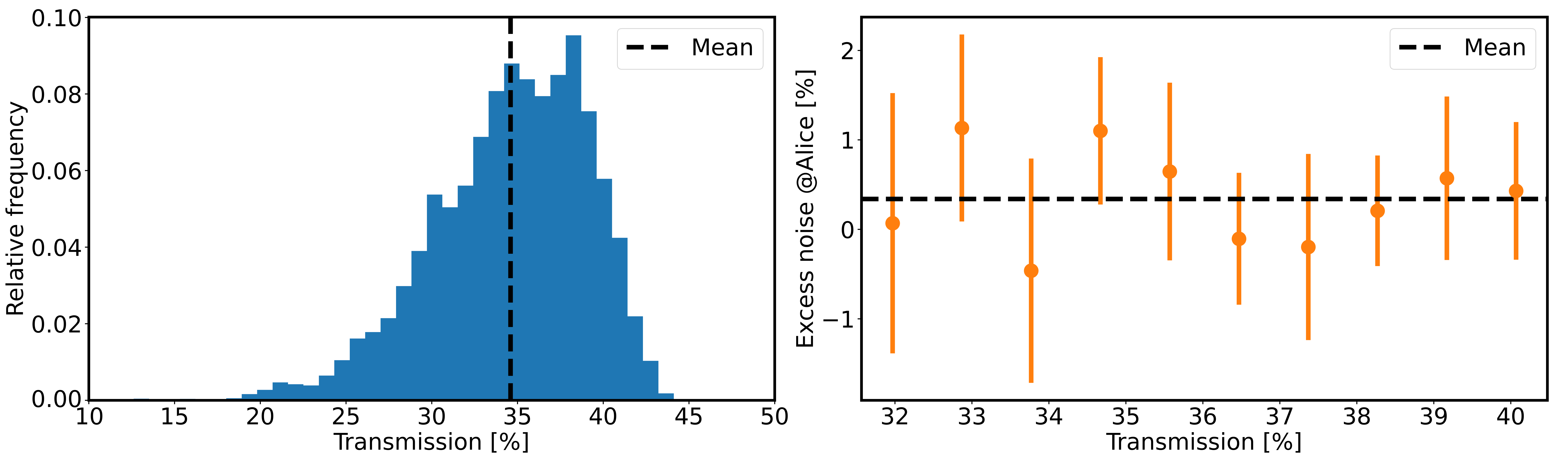}
    \caption{\label{fig:CV810results} Measurement with the free-space CV-QKD-810 setup in twilight conditions.
    Left: Transmission histogram of the fading channel, divided into sub-channels with 0.9\,\% width for further analysis.
    Right: Parameter estimation for the excess noise on the ten most populated sub-channels.
    The mean excess noise, defined as the weighted average according to the population size of the different sub-channels, was found as $0.34 \pm 0.31 \%$.
    }
\end{figure}

The compatibility of the CV-QKD-810 system with the overall network was tested in a point-to-point configuration. The integration of the system into the QKD network, including the optical interfaces to the TFT-1 and the electrical interfaces for classical radio frequency (RF) communications were successfully demonstrated. In this configuration, the system was able to successfully transmit the quantum states over the 1660\,m atmospheric channel between QuBUS and STW (FSO-Link-1) with a mean transmission of 34.6\,\% in twilight conditions (see Figure \ref{fig:CV810results}). After sub-binning the fluctuating channel into fixed transmission channels \cite{usenkoFadingChannel2012}, the measured parameters forecast positive key rates compatible with the newest security proofs for discrete modulated CV-QKD for both night and daylight operation
\cite{kanitscharFiniteSizeSecurityDiscreteModulated2023,jakschDMCVQKD}.

\subsection{\label{sec:experiment:fiber-cv-qkd}Direct link: Fiber \texorpdfstring{1550\,nm}{1550 nm} CV-QKD}

The QKD System CV-QKD-1550 was installed at the endpoints of the QuBUS-ACP fiber link (FIBER-Link-1, see Figure \ref{fig:geography}) with an overall transmission of ~98\%. 
As this system is a successor of the prototype originally developed for the QuNET demo in 2021, cf. App.~\ref{app:bonn-demo}, the campaign in Jena marked the first full-scale field test of this new iteration. A sucessful integration of the system into the overall network architecture was completed and a fully remote system control and operation over a time span of several days was showcased.

\begin{figure}[!t]
    \centering
    \includegraphics[width=0.8\textwidth]{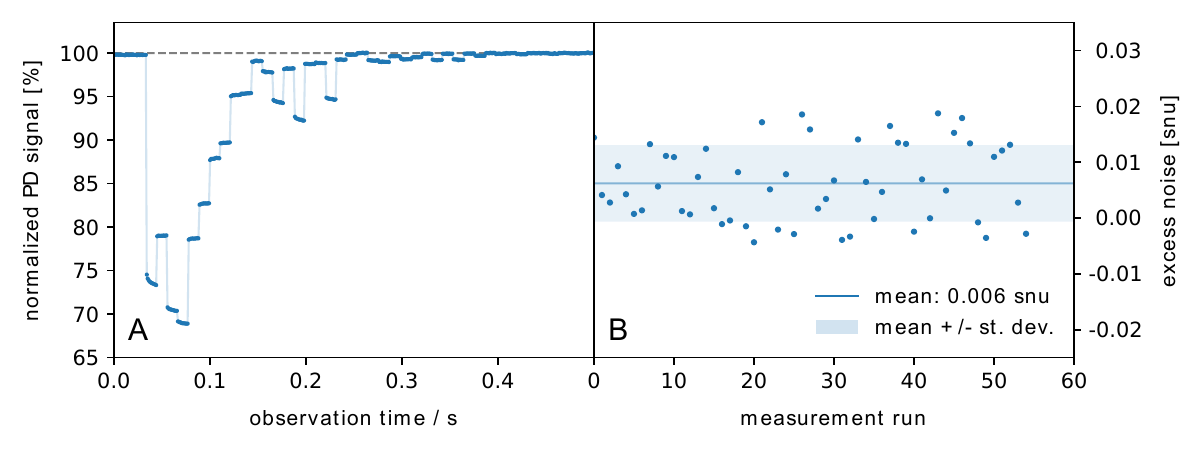}
    \caption{
        Measurement results for the fiber CV-QKD-1550 system.
        (a) Polarization control based on feedback from a single photodiode, reacting to a strong impulse disturbance.
        (b) Excess noise values for a subset of measurements taken during the campaign, with an average of $\xi = 6 \times 10^{-3} \, \mathrm{snu}$.
        }
    \label{fig:fiber-cv-qkd-plots}
\end{figure}

During this time, multiple measurement runs could even be performed autonomously over hours, without requiring any operator intervention. This was possible due to newly developed periodic recalibration and polarization control procedures. Even under strong turbulence conditions the system performed well and allowed extended exchanges of QPSK-modulated quantum states near the shot noise limit (see Figure \ref{fig:fiber-cv-qkd-plots}).

For a selected subset of these exchanges, a mean excess noise of $\xi = 6\cdot 10^{-3}\mathrm{\,snu}$ (shot noise units) was determined. 
Following an asymptotic security analysis as in \cite{zhang_improving_2012}, this would correspond to a secret key fraction of approx. $1.1 \times 10^{-2}\mathrm{\,bits/symbol}$, assuming the use of a rate $0.02$ Low Density Parity Check (LDPC) code available to us for error correction. While complete end-to-end key generation in the field remains challenging
for DM-CV-QKD systems, these findings mark a significant step towards this goal.
The analysis of the extensive data obtained during this campaign shows that positive asymptotic key rates can be achieved using QPSK modulation combined with realistic 
error correction implementations.

\subsection{\label{sec:experiment:eb-qkd}Direct link: Entanglement-based QKD}

\begin{figure}
    \centering
    \includegraphics[width=78.664mm]{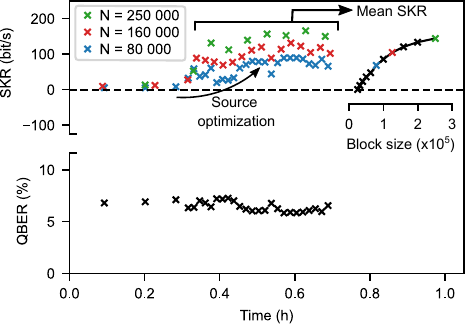}
    \caption{\label{fig:iof_QKD_SE1_Entanglement}
    Measurement results for the BBM92-QKD system.
    The finite-size secure-key rate (SKR) and quantum bit error rate (QBER) are shown for different block size sizes $N$. The block size of 160\,000 corresponds to a time for key generation of approximately 2\,min. When increasing the block size in the key post processing, the SKR approaches the asymptotic limit of 180\,bit/s.
    }
\end{figure}

The IOF entanglement-based system BBM92-QKD-IOF was deployed in ACP and generated an average secure-key rate of up to 130\,bit/s, considering 2-min blocks with size $N \approx 160\,000$ (Figure~\ref{fig:iof_QKD_SE1_Entanglement}) within the finite key analysis.
The photon pair source was located in the same building as the first receiver Alice (Figure~\ref{fig:iof_QKD_SE1_Entanglement_setup}).
The other receiver Bob was located in a nearby building and connected via a 1\,500 m fiber link with an overall transmission efficiency of 79\,\% (FIBER-Link-2).

The secret keys were generated considering finite-key-size effects and pushed into the GKMS.
The finite-key length $l$ was calculated according to publications \cite{krzicMetropolitanFreespaceQuantum2023,tomamichelTightFiniteKeyAnalysis2012}.
The overall security parameter $\epsilon$ was picked as $10^{-10}$ and the tolerated quantum bit error rate $Q_\mathrm{tol}$ set to 10\% after link characterization.
Blocks with QBERs exceeding 10\% in the parameter estimation step were discarded.

\subsection{\label{sec:experiment:fwf-dv-qkd}Direct link: Fiber-wireless-fiber DV-QKD}

FSO links provide a viable alternative for situations were a direct fiber-optic connection between two locations is unattainable.
However, the necessity for a clear line of sight between telescopes imposes limitations on their placement. Consequently, telescopes are frequently installed on rooftops. In contrast, QKD systems require a secure indoor installation to guard against tampering. This conflict is resolved through FWF links, where indoor QKD systems are connected to the external telescopes on both ends.
The BB84-QKD system from Fraunhofer HHI was continuously operated for 34\,h during day and night over an FWF link between the two main areas of the Jena QKD testbed.
During this time the system achieved a QBER of $1.78 \, \mathrm{\%}$ ($2.67 \, \mathrm{\%}$) for the $\mathsf{Z}$-basis ($\mathsf{X}$-basis) and produced $53.0\,\mathrm{Mbit}$ of key at an average SKR of $433\,\mathrm{bit/s}$.

For the combined channel (FIBER-Link-3 + FSO-Link-1 + FIBER-Link-1, cf.~Table\,\ref{tab:Subsystems1}) an average transmission of
$-20.3 \pm 0.9 \, \mathrm{dB}$
and $C_n^2$ scintillation values ranging from $5.86 \times 10^{-16}\,\mathrm{m^{-2/3}}$ to $6.84 \times 10^{-14}\,\mathrm{m^{-2/3}}$, with solar irradiance peaking at $944\,\mathrm{W/m^2}$ were measured.

\subsection{\label{sec:experiment:trusted-node}Trusted-node operation}

\begin{figure}
    \centering
    \includegraphics[width=\linewidth]{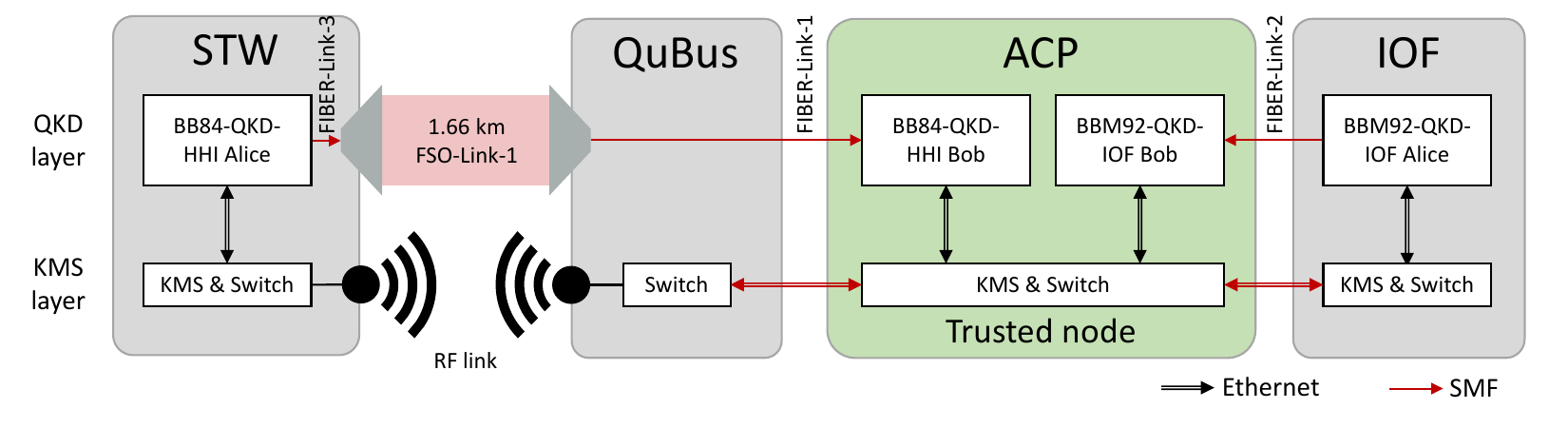}
    \caption{\label{fig:trusted-node-schematic}
    Trusted node link configuration.
    The HHI timebin-phase BB84-QKD system was operated over the fiber-wireless-fiber (FWF) link between STW and ACP, the IOF BBM92-QKD system was operated over fiber between ACP and IOF.
    RF: Radio-frequency;
    KMS: Key management system.
    }
\end{figure}

Direct links between two communicating parties are not always available such that intermediate trusted nodes have to be used.
In this scenario, two different QKD links were used in a point-to-point configuration with an intermediate trusted relay node at ACP, see Figure\,\ref{fig:trusted-node-schematic}.

The first QKD link was established over a FWF link realized by the BB84-QKD system, cf. Sec. \ref{sec:subsystems:qkd:hhi-1550-bb84}. The link consisted of three different segments: 20~m fiber (FIBER-Link-3), 1660~m free space (FSO-Link-1), and 300~m fiber (FIBER-Link-1). The BB84-QKD system achieved an average QBER of $1.75\,\%$ ($3.13\,\%$) for the $\mathsf{Z}$-basis ($\mathsf{X}$-basis), leading to an average SKR of $385\,\mathrm{bit/s}$ over the FWF link during daylight.  
The second QKD link was realized over a 300\,m fiber channel between ACP and IOF (FIBER-Link-2) and utilized the BBM92-QKD system, cf. Sec. \ref{sec:subsystems:qkd:iof-1550-bbm92}. On this link an average SKR of 120\,bit/s with an average QBER of 6\,\% using a block size of 160 000 was achieved, equivalent to retrieving new secure key material every 2\,min.
Both links were operated simultaneously for 3.5~h during daytime in a trusted-node scenario. For the FWF link an average transmission of $-20.9 \pm 0.5 \, \mathrm{dB}$, an average scintillation of $C_n^2 = 1.85 \times 10^{-14}\,\mathrm{m^{-2/3}}$, and an average solar irradiance of $591\,\mathrm{W/m^2}$ were measured. 
In order to obtain symmetric cryptographic keys at the STW and IOF sites, the keys generated by the two systems via different link types were logically combined by the key management layer, cf. Sec.~\ref{sec:architecture}.
As a typical user application, quantum-secure access from the IOF-LAN to a Nextcloud server, offering file storage and collaborative editing functionality, hosted in the STW-LAN was demonstrated.

\section{\label{sec:discussion}Conclusion}

An ad-hoc QKD network was demonstrated, utilizing both free-space and fiber quantum channels. Compact, transportable 200 mm Free-Space Optical (FSO) terminals were employed, enabling the swift setup and facilitating real-time QKD operations during both day and night over distances approximately up to 2 km.
Interoperability of a multitude of QKD systems was successfully established, encompassing a fiber-optimized CV-QKD system at 1550 nm, a free-space-optimized CV-QKD system at 810 nm, a 1-decoy time-bin BB84 system at 1550 nm, a high-dimensional coding prepare-and-measure system, and an entanglement-based BBM92 system. Using these systems, multiple QKD links were established via direct fiber and multi-kilometer free-space links.
A Fixed Wireless Fiber (FWF) link was demonstrated, which permitted the placement of QKD systems in secure rooms inside buildings without limiting the free-space terminal placement. A trusted-node link was also exhibited, showing the potential for key generation across different system domains and its application for encrypted remote file server access.
Overall, these results substantiate the interoperability of various QKD systems and the adaptability of a rapidly established ad-hoc QKD network. Despite the transitory nature of the Jena testbed, its accessibility has been of immense value for the ongoing advancement of quantum communication systems and their subsystems across all project partners. Detailed descriptions of each individual experiments, however, will be published by the respective project partners.

\section{Acknowledgements}

This research was conducted within the scope of the QuNET Initiative,
funded by the German Federal Ministry of Education and Research (BMBF)
in the context of the federal government's research framework in
IT-security “Digital. Secure. Sovereign.”. The research was supported by the provision of important infrastructure by Stadtwerke Jena GmbH. We would like to express our sincere thanks for this ongoing support.

\section{\label{sec:author_contributions_jan}Author contributions}

M.G., J.K. and Ö.B. contributed equally to this work.
M.G., J.K., Ö.B., T.D., F.F., R.F., K.J., A.K., S.M., F.M., K.P., R.R., G.S., M.S., C.S., F.S., A.T., and N.W. conceptualized the fundamental ideas and supervised the project's implementation.
M.G., J.K., Ö.B., P.A., T.D., N.D., J.D., J.H., K.J., T.K., A.K., O.K., M.L., K.L.M., S.M., N.P., K.P., S.R., M.R., R.R., G.S., M.S., J.S., A.S., S.S., S.Sp., C.S., F.S., H.V., and N.W. conducted the experiments and analyzed the data.
M.G., Ö.B., F.D., F.F., R.F., T.G., G.L., C.M., A.M., F.M., N.P., M.S., F.S., A.T., H.V., and N.W. handled funding and administration.
M.G., J.K., Ö.B., P.A., F.D., T.D., N.D., J.D., R.F., F.F., T.G., J.H., K.J., O.K., T.K., A.K., M.L., G.L., C.M., K.L.M., A.M., S.M., F.M., K.P., N.P., S.R., M.R., R.R., G.S., M.S., J.S., A.S., S.S., S.Sp., C.S., F.S., A.T., H.V., N.W., and S.W. wrote the manuscript.
All authors discussed and revised the manuscript.

\begin{appendices}

\section{\label{app:bonn-demo}Quantum Key Distribution between Federal Agencies}

In the scope of the German national initiative QuNET
\cite{QuNETInitiative2019},
a first QKD demonstration experiment was performed in 2021 with the main goal of demonstrating technical capabilities and the interoperability of QKD systems employing different protocols in a heterogeneous QKD network.
Therefore, the German Federal Ministry of Education and Research (Bundesministerium für Bildung und Forschung, BMBF) and the German Federal Office for Information Security (Bundesamt für Sicherheit in der Informationstechnik, BSI) were connected via a fiber link and a free-space optical link, both of which were used for both quantum and classical channels.

\begin{figure}
    \centering
    \includegraphics[width=0.9\columnwidth]{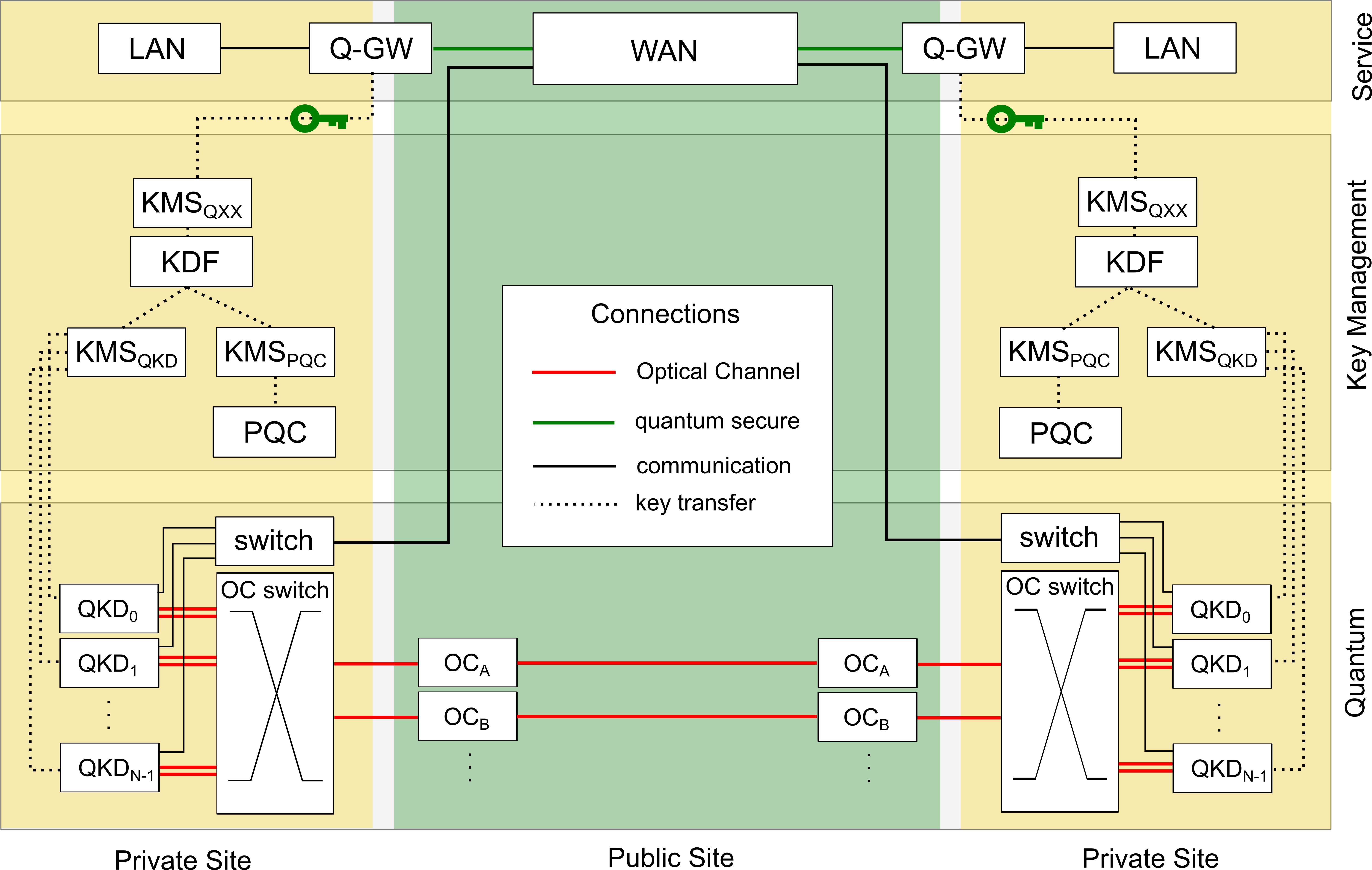}
    \caption{
    System Architecture of the Bonn demo experiment.
    Four QKD systems generated key material (quantum layer, top), which were cryptograhically combined with keys from a post-quantum cipher in the key management (KMS) layer (middle).
    The application layer (bottom) transparently handled the encryption via quantum-secure gateways. 
    OC: Optical channel;
    PQC: Post-quantum cryptography;
    KMS: Key management system;
    KDF: Key derivation function;
    LAN: Local area network;
    WAN: Wide area network;
    Q-GW: Quantum-secure gateway (encryptor).
    }
    \label{fig:int_SysArchitecture}
\end{figure}

The used system architecture for a point-to-point network architecture, cf. Figure~\ref{fig:int_SysArchitecture}, provided the basis for the architecture used in the Jena testbed, cf. Figure~\ref{fig:qkd-i-3}.
In addition, to mitigate potential attacks against individual QKD systems, it also allowed for the cryptographic combination of keys from multiple QKD systems and a post-quantum cryptography (PQC) algorithm.

Over the course of the measurement campaign, four QKD demonstrators were operated: A discretely modulated CV-QKD system operating at 810\,nm, an entanglement-based DV-QKD system operating at 810\,nm, a discretely modulated CV-QKD system operating via fiber in the C-band,
and a timebin-phase BB84 DV-QKD system operating via fiber and an FSO link in the C-band.

As Q-GW, Layer-2 encryptors (R\&S\textsuperscript{\textregistered} SITLine ETH40G with clearance for VS-NfDEU RESTRICTED \& NATO RESTRICTED by BSI) with 40\,Gbit/s throughput were employed and pulled a 256 bit key every two minutes.
Furthermore, the system resilience against quantum channel interruptions or QKD system maintenance was increased by a buffered KMS operation.

As a proof-of-concept application a video conference call between the two private sites BMBF and BSI was encrypted using a key derived from multiple QKD systems and a PQC algorithm.

\end{appendices}

\section*{References}
\bibliographystyle{unsrt}
\bibliography{literature}

\end{document}